\documentclass[usenatbib,usegraphicx]{mn2e}
\usepackage{amsmath}
\usepackage{amsfonts}
\usepackage{amssymb}

\title[The UV properties of E+A galaxies]
    {The UV properties of E+A galaxies: constraints on feedback-driven quenching of star formation}

\author[S.~Kaviraj et al.]
{S. Kaviraj\thanks{E-mail: skaviraj@astro.ox.ac.uk}$^{1}$, L. A. Kirkby$^{1}$, J. Silk$^{1}$ and M. Sarzi$^{2}$\\
$^1$Department of Physics, University of Oxford, Keble Road,
  Oxford OX1 3RH, UK\\
$^2$Centre for Astrophysics Research, University of Hertfordshire,
Hatfield, AL10 9AB, UK}

\begin{document}

\date{29 May 2007}

\pagerange{\pageref{firstpage}--\pageref{lastpage}} \pubyear{2007}

\maketitle

\label{firstpage}


\begin{abstract}
We present the first large-scale study of E+A galaxies that
incorporates photometry in the ultraviolet ($UV$) wavelengths. E+A
galaxies are `post-starburst' systems, with strong Balmer
absorption lines indicating significant recent star formation, but
without [OII] and H$\alpha$ emission lines which are
characteristic of ongoing star formation. The starburst that
creates the E+A galaxy typically takes place within the last Gyr
and creates a high fraction (20-60 percent) of the stellar mass in
the remnant over a short timescale ($<0.1$ Gyrs). We find a tight
correlation between the luminosity of our E+A galaxies and the
implied star formation rate (SFR) during the starburst. While
low-luminosity E+As ($M(z)>-20$) exhibit implied SFRs of less than
50 $M_{\odot} yr^{-1}$, their luminous counterparts ($M(z)<-22$)
shows SFRs greater than 300 and as high as 2000 $M_{\odot}
yr^{-1}$, suggesting that luminous and ultra-luminous infrared
galaxies in the low-redshift Universe could be the progenitors of
massive nearby E+A galaxies. We perform a comprehensive study of
the characteristics of the quenching that truncates the starburst
in E+A systems. We find that, for galaxies less massive than
$10^{10}M_{\odot}$, the \emph{quenching efficiency} decreases as
the galaxy mass increases. However, for galaxies more massive than
$10^{10}M_{\odot}$, this trend is reversed and the quenching
efficiency increases with galaxy mass. Noting that the mass
threshold at which this reversal occurs is in excellent agreement
with the mass above which AGN become significantly more abundant
in nearby galaxies, we use simple energetic arguments to show that
the bimodal behaviour of the quenching efficiency is consistent
with AGN and supernovae (SN) being the principal sources of
negative feedback above and below $M\sim10^{10}M_{\odot}$
respectively. The arguments assume that quenching occurs through
the mechanical ejection or dispersal of the gas reservoir and
that, in the high mass regime ($M>10^{10}M_{\odot}$), the
Eddington ratios in this sample of galaxies scale as $M^{\gamma}$,
where $1<\gamma<3$. Finally, we use our E+A sample to estimate the
time it takes for galaxies to migrate from the blue cloud to the
red sequence. We find migration times between 1 and 5 Gyrs, with a
median value of 1.5 Gyrs.
\end{abstract}


\begin{keywords}
galaxies: elliptical and lenticular, cD -- galaxies: evolution --
galaxies: formation -- galaxies: fundamental parameters
\end{keywords}


\section{Introduction}
`E+A' galaxies exhibit strong Balmer absorption lines,
characteristic of significant recent star formation, but lack [OII]
or H$\alpha$ emission which are characteristic of ongoing star
formation. Their spectral features indicate that a vigorous episode of recent star
formation in these systems has been quenched abruptly. As
`post-starburst' systems, E+A galaxies provide a valuable
evolutionary link between the gas-rich star-forming population and
gas-poor quiescent systems. The persistent dichotomy in galaxy
colours over a large range in redshift, coupled with the gradual
build-up of the red sequence over time \citep[e.g.][]{Bell2004}, indicates that \emph{how}
and \emph{over what timescale} galaxies evolve from the `blue
cloud' to the `red sequence' is of significant importance in
understanding the macrosopic evolution of the galaxy population
over the lifetime of the Universe. The characteristics of E+A
galaxies provide a unique (but transient!) window into the very
point where this transition from blue-cloud to red sequence is
about to begin. Understanding what drives their vigorous star
formation episode and, in particular, what causes it to be
quenched so suddenly is clearly an important step in our
understanding of galaxy evolution.

First noticed by \citet{Dressler1983} in distant ($z\sim0.4$)
clusters, E+A galaxies have been subsequently found to be abundant in all types
of environments. At intermediate redshifts ($0.3<z<1$), the
proportion of E+A galaxies appears to be a factor of four greater
in clusters than in the field \citep{Tran2004}. In the nearby
Universe E+A galaxies have been detected mainly in the field
\citep{Zabludoff1996,Quintero2004} primarily because most galaxies
do not reside in dense cluster-like environments. Indeed, the
overall distribution of the local environments of E+A systems
follows that of the general galaxy population as a whole (Blake et
al. 2004). The fraction of E+A systems in clusters shows a rapid
decline from intermediate redshifts ($z\sim0.5$) where it is
typically higher than 20 percent
\citep[e.g.][]{Couch1987,Belloni1995} to less than 1 percent in
local clusters \citep{Fabricant1991}.

Several plausible explanations for the E+A phenomenon have been
proposed and studies indicate that there are multiple channels for
creating such post-starburst spectral signatures. Since E+A galaxies were
first detected in clusters, it was initially thought that their
production required a cluster-specific mechanism, such as galaxy
harassment or ram-pressure stripping. However, their abundant
presence in the field indicates that other channels exist for the
production of E+A systems \citep[e.g.][]{Goto2005a}, although
cluster-specific mechanisms may contribute or even dominate their
evolutionary pathways in dense regions of the Universe.

Another possible explanation for E+A-like spectra is that the
optical emission lines are heavily suppressed by dust obscuration
\citep[e.g.][]{Smail1999,Couch1987,Poggianti2000}. An efficient
test of this scenario is to observe the E+A population in the
radio wavelengths where the effects of dust obscuration are
absent. However, a radio study of 15 galaxies in the
\citet{Zabludoff1996} sample by \citet{Miller2001} detected only
two at radio luminosities which are consistent with modest levels
of star formation. Similarly, \citet{Goto2004} performed VLA 20 cm
radio-continuum observations of 36 E+A galaxies drawn from the
SDSS and found that none of these galaxies were detected in the
radio wavelengths. These results are supported by
\citet{Galaz2000} who do not find compelling evidence for strong
internal dust extinction in their E+A sample.

Some E+A galaxies exhibit morphological disturbances indicative of
interactions with near neighbours. \citet{Yang2004} have studied the
morphologies of the five bluest E+A galaxies in the
sample of \citet{Zabludoff1996} using the WFPC2 camera on board the \emph{Hubble Space Telescope
  (HST)} and found that four out of these five galaxies
display morphological disturbances consistent with recent mergers.
Similarly, \citet{Goto2004} have shown, using the largest sample
of E+A galaxies studied to date, that these systems have an excess
of local galaxy density at spatial scales less than 100 kpc but
not at scales spanned by galaxy clusters ($\sim1$ Mpc) or
large-scale structure ($\sim8$ Mpc). They find that $\sim30$
percent of E+A galaxies exhibit morphological disturbances,
indicating that their evolution is linked, at least partially, to
mergers and interactions. It is worth noting that, although the
\citet{Goto2004} sample is larger and their results are
statistically more robust, detection of small, low-surface
brightness (tidal) features probably requires both deeper and
higher resolution imaging than that provided by the SDSS. Hence, the
fraction of E+A galaxies with tidal tails may well be higher than
$\sim30$ percent in the \citet{Goto2004} sample, making the
conclusions from these two studies more consistent.

From a theoretical perspective, the merger hypothesis is supported by
numerical simulations which indicate that gas-rich mergers are capable
of triggering strong star formation episodes. The immediate aftermath of such an
event is predicted to be a largely spheroidal remnant with a blue
core, widespread morphological disturbances such as tidal tails and
young star clusters \citep[e.g.][]{Mihos1994,Bekki2001}. Indeed, the
study of \citet{Yang2004} has found a striking correspondence between
the predicted properties of merger remnants and the observed
characteristics of blue E+A galaxies.

Numerous authors have studied E+A galaxies using a variety of
indicators, which allows us to infer the characteristics of the
(recent) starburst that dominates the spectra of these systems.
While the age of the burst should be approximately $\sim1$ Gyr,
simply by virtue of A-type stars being present, the mass fractions
forming in the burst are inferred to be quite high.
\citet{Yang2004} estimate that 50 percent of the mass may form in
the burst, allowing the E+A galaxies to fade on to the E/S0
fundamental plane in a few Gyrs after the event.
\citet{Norton2001}, who analysed the entire \citet{Zabludoff1996}
sample, estimate that the fraction of stellar mass in young
(A-type) stars is typically between 30 and 80 percent.
\citet{Bressan2001} infer a similarly high mass fraction of 60
percent for the sample of E+A galaxies in \citet{Poggianti2001},
while \citet{Liu1996} derive mass fractions typically greater than
20 percent (and as high as 100 percent) for eight E+A galaxies
taken from the literature (but note that \citet{Liu2007} revise
the inferred mass fraction to lower values for one of their E+A
galaxies - G515). While E+A's are, by definition, presently
quiescent, an estimate of the star formation rate (SFR) that
\emph{might} have characterised the recent starburst can be
estimated from dusty starburst galaxies which show an E+A
signature in their spectra. \citet{Poggianti2000} estimate the
SFRs in these systems to be between 50 and 300 M${\odot}$ per
year.

While mergers and interactions might trigger star formation, the
exact mechanism by which the starburst is subsequently quenched
remains largely unexplained. Star formation is, of course,
inherently subject to `negative feedback', since the star
formation rate (SFR) depends on the mass of cold gas available. As
star formation proceeds and the gas reservoir is exhausted, the
rate of further star formation activity correspondingly decreases.
However, the properties of E+A galaxies are generally inconsistent
with such a gradual, self-modulated decline in the SFR and imply a
more abrupt quenching of star formation activity.

\begin{figure}
\begin{center}
\includegraphics[width=3.5in]{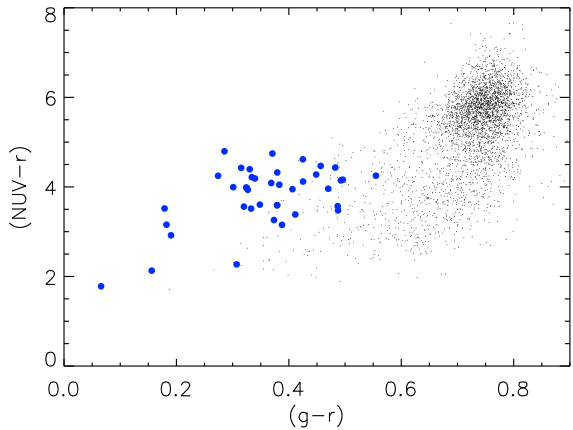}
\caption{The position of our E+A galaxies (filled blue circles) in
the $(NUV-r)$ vs (g-r) colour space, compared to a sample of
early-type galaxies drawn from the SDSS DR5 using the method of
Kaviraj et al. (2006b) and cross-matched with GALEX. The E+A
population inhabits the blue cloud and is well separated from the
early-type galaxies which define the bulk of the red sequence.
Note that the colours are in the observed frame i.e. they are
shown without applying K-corrections.} \label{fig:cmspace}
\end{center}
\end{figure}

Common sources of kinetic and thermal `feedback' that might be
capable of quenching star formation, e.g. by ejecting or heating
the gas reservoir, include supernovae (SN), which are effective
mainly in low-mass galaxies due to their shallow potential wells,
and AGN, which are thought to operate in massive galaxies where SN
feedback is not energetic enough to `disable' the cold gas
reservoir. \citet{Goto2006} have demonstrated that some galaxies
which contain signatures of an active AGN also exhibit
post-starburst signatures, indicating that a connection exists
between these two phenomena. They find that the post-starburst
regions are centred around the AGN, the spatial proximity implying
that AGN activity could plausibly affect the region around it.
Indeed, the fact that the AGN \emph{outlives} the starburst is an
indication that it may have played a role in the quenching of star
formation.

In this study we extend the existing literature on post-starburst
galaxies by performing the first study of E+A systems that
incorporates their ultraviolet ($UV$) photometry. The extreme
sensitivity of the $UV$ to young stars allows us to put strong
constraints on the star formation history (SFH) of galaxies within
the last $\sim2$ Gyrs. The addition of $UV$ photometry alleviates
the age-metallicity degeneracy that commonly plagues optical
studies \citep{Kaviraj2007}, allowing us to better quantify the
recent SFH than can be achieved using optical data alone. Indeed,
the sensitivity of the $UV$ to young stars has recently been been
exploited to demonstrate that early-type galaxies, traditionally
thought to be entirely quiescent systems, exhibit
\emph{widespread} evidence for low-level recent star formation,
consistent with the expectations of the standard LCDM model
(Kaviraj et al. 2006a).

In this paper, we exploit the properties of the $UV$ to accurately
reconstruct the recent star formation history of individual E+A
galaxies. By comparison to a large library of models we derive
estimates for the ages of the recent bursts, the mass fractions
formed in them and the timescales over which they occurred. By
comparing the derived timescales to the dynamical timescales of
the galaxies, we explore the efficiency with which star formation
has been quenched in these systems and study how this quenching
efficiency varies as a function of galaxy properties such as mass,
luminosity and stellar velocity dispersion. We use simple
energetic arguments to study whether the characteristics of the
quenching are consistent with feedback from supernovae and AGN.
Finally, we study the timescale over which E+A galaxies can evolve
from their present positions in the blue cloud to the red
sequence. Since E+As are, by definition, at the point of
truncation of their star formation, it is possible to get a rather
`clean' estimate of this migration timescale by studying the
evolution in both $UV$ and optical colours.

\begin{figure}
\begin{center}
\includegraphics[width=3in]{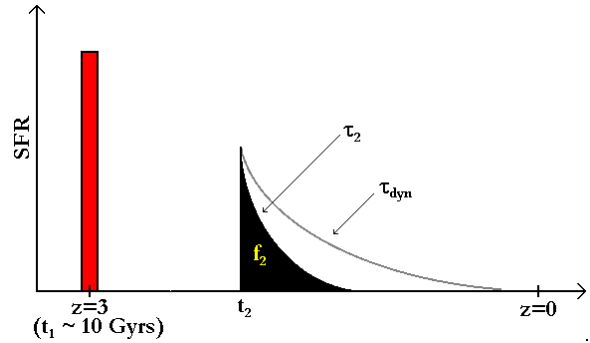}
\caption{Model SFHs (see Section 2) are constructed by assuming
that an instantaneous burst of star formation at high redshift
($z\sim3$) is followed by a second burst, modelled as an
exponential. The free parameters are the age ($t_2$), mass
fraction ($f_2$) and timescale ($\tau_2$) of the second burst.
$t_2$ is allowed to vary from 0.05 Gyrs to the look-back time
corresponding to $z=3$ in the rest-frame of each E+A galaxy. $f_2$
varies between 0 and 1 and $\tau_2$ is allowed to vary from 0.01
Gyrs to 4 Gyrs. $\tau_{dyn}$ indicates the dynamical timescale of
the galaxy. More catastrophic quenching leads to smaller values of
the ratio $\tau_2/\tau_{dyn}$, which is used in Section 5 to
parametrise the \emph{efficiency} with which quenching occurs in
the E+A galaxies studied here.}\label{fig:sfh_cartoon}
\end{center}
\end{figure}


\section{Sample selection}
We use the publicly available Garching SDSS
catalog\footnote{http://www.mpa-garching.mpg.de/SDSS/DR4/} to
select E+A galaxies from the SDSS DR4. The criteria used for E+A
selection follow that of \citet{Goto2005a}: H$\delta$ (EW) $>
5.0\AA$ (measured in the wider window of 4082-4122$\AA$),
H$\alpha$ (EW) $> -3.0\AA$, [OII] (EW) $> -2.5\AA$. Absorption
lines have a positive sign in these definitions. Note that
including the equivalent width of the H$\alpha$ line makes the
selection criteria stronger, because using \emph{only} the [OII]
line could result in more than 50 percent of the selected galaxies
having measurable H$\alpha$ emission (e.g. Goto et al. 2003). In
addition to the line measurements we also use estimates of stellar
mass and AGN classifications provided by the Garching SDSS
catalog. The line index measurements in this catalog were
calculated using the code of \citet{Tremonti2004}, while stellar
mass estimates and AGN classifications were taken from a series of
comprehensive analyses of the local galaxy population observed by
the SDSS (Kauffmann et al. 2003a; Kauffmann et al. 2003b;
Brinchmann et al. 2004)

To ensure the accuracy of the line measurements, the sample was
restricted to galaxies where the median S/N of the spectrum is
greater than 10. Galaxies which show evidence of an active AGN
were removed, because scattered light from the AGN could
contaminate the $UV$ continuum and affect the derived parameters
presented later in this analysis. Finally, the sample was
restricted to galaxies in the redshift range $0<z<0.2$, since the
$UV$ filters used in the analysis (see below) trace flux blueward
of the $UV$ continuum at higher redshifts.

This E+A sample, drawn from the SDSS DR4, was then cross-matched
with publicly available $UV$ photometry from the second data
release of the GALEX mission (Martin et al. 2005). GALEX provides
two $UV$ filters: the far-ultraviolet ($FUV$), centred at
$\sim1530\AA$ and the near-ultraviolet ($NUV$), centred at
$\sim2310\AA$. This cross-matching produced
38 E+A galaxies which
have \emph{at least} a detection in the $NUV$ filter, 28 of which
have photometry in both the $FUV$ and $NUV$ filters.

Figure \ref{fig:cmspace} shows the position of our E+A galaxies
(filled blue circles) in the $(NUV-r)$ vs $(g-r)$ colour space,
compared to a sample of early-type galaxies drawn from the SDSS
DR5 using the method of Kaviraj et al. (2006b) and cross-matched
with the GALEX second data release. It is apparent that the E+A
population is well separated from the early-type population which
defines the bulk of the red sequence. Comparison to Figure 1 in Yi
et al. (2005) indicates that E+A galaxies inhabit the blue cloud.

Since previous studies have indicated that E+A galaxies are
potential progenitors of early-type galaxies, we check the
morphologies of the E+A galaxies in our sample using the SDSS
$fracDev$ parameter. The SDSS pipeline fits both deVaucouleur's
and exponential surface brightness profiles to galaxy images (in
each filter) and creates a composite \emph{best-fit} profile using
a linear combination of the two fits. The $fracDev$ parameter is
the weight of the deVaucouleur's profile in this composite fit.
Thus, galaxies with large values of $fracDev$ can be considered to
have spheroidal morphology. We find that all but 6 of the galaxies
studied in this sample have $fracDev>0.8$ in the $r$-band, which
indicates that the majority of E+A galaxies in this sample are
indeed spheroidal systems. It is worth noting that all 6 galaxies
which have $fracDev<0.8$ are low-mass systems
($M<10^{10}M_{\odot}$) and their non-deVaucouleur's morphologies
are consistent with the fact that dwarf ellipticals typically show
exponential surface brightness profiles.


\section{Parameter estimation}
We estimate parameters governing the star formation history (SFH)
of each E+A galaxy by comparing their $(FUV, NUV, u, g, r, i, z)$
photometry to a library of synthetic photometry generated using a
large collection of model SFHs. Each SFH is constructed by
assuming that an instantaneous burst of star formation at high
redshift ($z\sim3$) is followed by a second burst which is
modelled as an exponential.

Figure \ref{fig:sfh_cartoon} shows a schematic representation of
the model SFHs. The free parameters in this analysis are the age
($t_2$), mass fraction ($f_2$) and timescale ($\tau_2$) of the
second burst. $t_2$ is allowed to vary from 0.05 Gyrs to the
look-back time corresponding to $z=3$ in the rest-frame of each
E+A galaxy. $f_2$ varies between 0 and 1 and $\tau_2$ is allowed
to vary from 0.01 Gyrs to 4 Gyrs. Since our focus is purely on E+A
galaxies, we must, by definition, exclude models which contain
ongoing star formation. Therefore, we only keep models where the
intensity of star formation in the second burst at present-day is
less than 5 percent of the intensity when the burst
started\footnote{This implies that models with small values of
$t_2/\tau_2$ are excluded.}. Changing this threshold to less than
5 percent leaves our results unchanged. Note that excluding models
that have ongoing star formation reduces the allowed parameter
space of model SFHs, resulting in much tighter constraints on the
parameters, in particular, the timescale of star formation
$\tau_2$.

The motivation for an instantaneous burst at high redshift stems
from the fact that, as potential progenitors of spheroidal
galaxies, one expects a substantial fraction of stellar mass to
have formed at high redshift in these galaxies
\citep[e.g.][Kaviraj et al. 2006b]{BLE92}. Furthermore, since the
$UV$ is insensitive to stellar populations older than $\sim2$ Gyrs
coupled with the fact that optical colour evolution virtually
stops after 5-6 Gyrs \citep[e.g.][]{Yi2003}, resolving this old
burst does not affect the $UV$ and optical colours of the model.

The $UV$ flux in each model is completely dominated by the second
burst of star formation. Even though our expectations, on the
basis of the properties of E+A galaxies (and previous studies of
E+A systems), are that $t_2 \sim1$ Gyr and $f_2$ is high, our
model library does not put a prior on these parameters - vigorous
starbursts at recent epochs are allowed, as is slowly declining
star formation from intermediate redshifts. In essence, the
comparison between the synthetic library and observed photometry
is allowed to `choose' the optimum SFH that satisfies the
photometry of each E+A galaxy.

To build the library of synthetic photometry, each model SFH is
combined with a single metallicity in the range 0.1Z$_{\odot}$ to
2.5Z$_{\odot}$ and a value of dust extinction parametrised by \emph{E(B-V)} in the range 0 to 0.5.
Photometric predictions are generated by combining each model SFH
with the chosen metallicity and \emph{E(B-V)} values and convolving with
the stellar models of \citet{Yi2003} through the GALEX $FUV$,
$NUV$ and SDSS $u, g, r, i, z$ filters. This procedure yields a
synthetic library of 1.5 million models.

Since our E+A galaxies are observed at a range of redshifts,
equivalent libraries are constructed at redshift intervals of
$\delta z=0.01$. A fine redshift grid is essential in such a low
redshift study because a small change in redshift produces a
relatively large change in look-back time over which the $UV$ flux
can change substantially, inducing `K-correction-like' errors into
the analysis.

The free parameters ($t_2$, $f_2$ and $\tau_2$) are estimated by
comparing each observed galaxy to every model in the synthetic
library, with the likelihood of each model ($\exp -\chi^2/2$)
calculated using the value of $\chi^2$ computed in the standard
way. From the joint probability distribution, each parameter is
marginalised to extract its one-dimensional probability density
function (PDF). We take the median of this PDF as the best
estimate of the parameter in question and the 16 and 84 percentile
values as the `one-sigma' uncertainties on this estimate. The
cosmological parameters used in this study assume a $\Lambda$CDM
model: $h=0.7$, $\Omega_{m}=0.3$ and $\Omega_{\Lambda}=0.7$.

Note that there are 10 galaxies in our sample which do not have
$FUV$ detections. The parameter estimation on these galaxies is
carried out using only the $NUV$, $u, g, r, i$ and $z$-band
filters. We do not use the $FUV$ detection limit as an upper limit
to the $FUV$ flux.

\begin{figure}
$\begin{array}{c}
\includegraphics[width=3.5in]{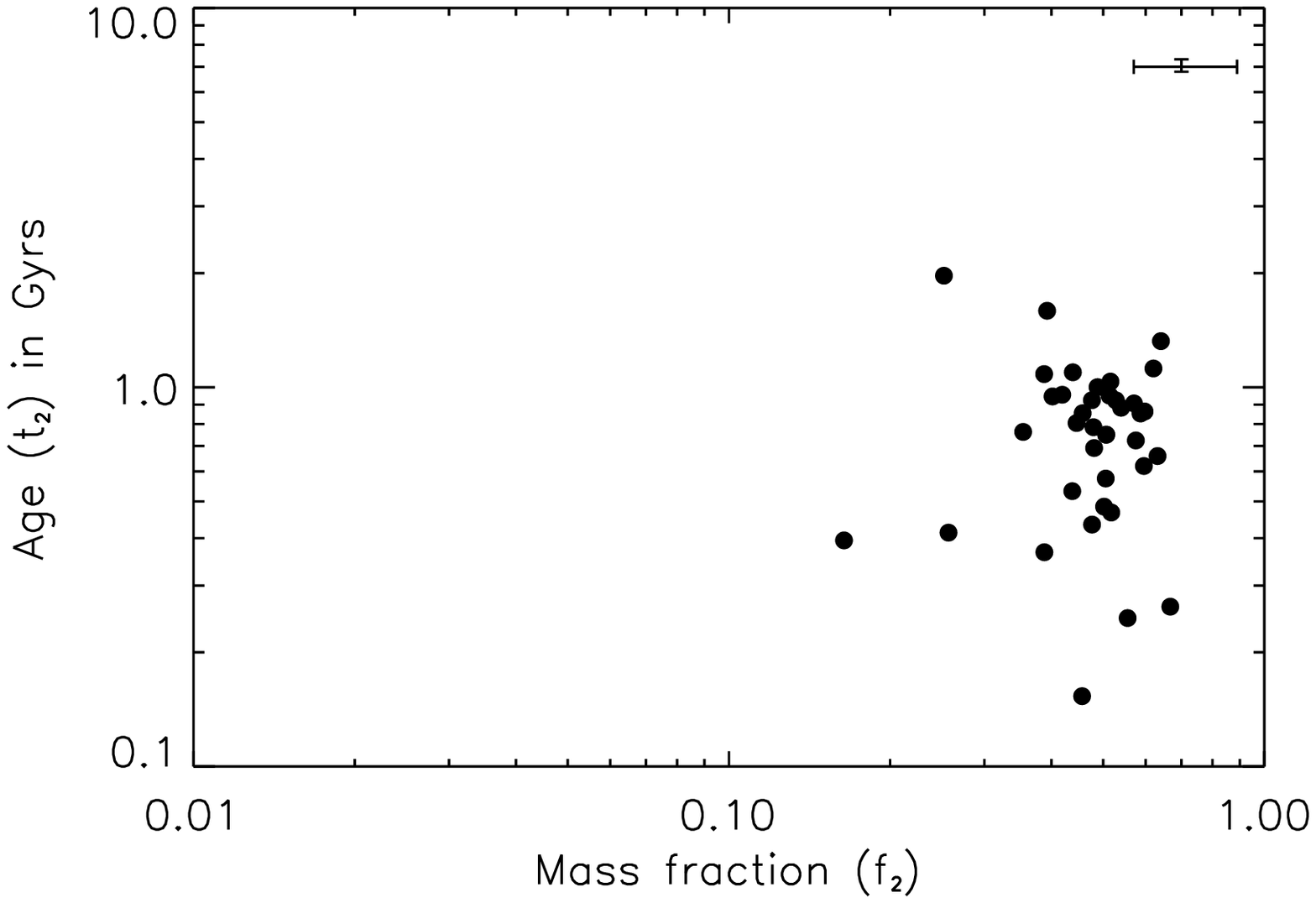}\\
\includegraphics[width=3.5in]{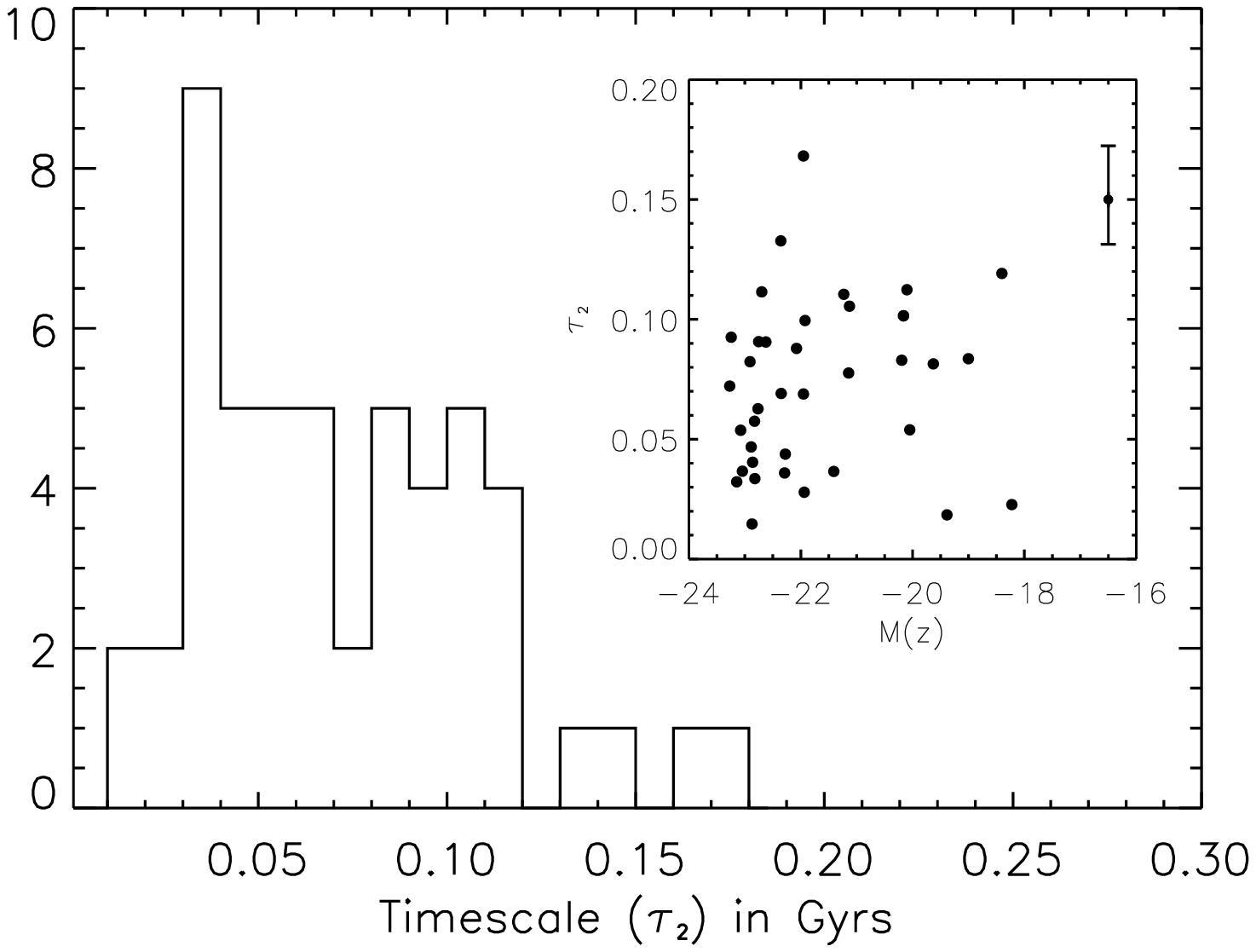}
\end{array}$
\caption{TOP PANEL: Ages and mass fractions of the burst that
creates
  the E+A galaxies in our sample. Burst ages are typically within a Gyr,
as one expects from the lifetimes of A-type stars which contribute
significantly to E+A spectra. We recover large mass fractions,
typically higher than 10 percent and possibly as high as 70
percent, consistent with previous studies. BOTTOM PANEL: Effective
timescales of the starbursts in E+A galaxies. The timescales are
typically short - between 0.01 and 0.2 Gyrs - implying that the
star formation rates during the burst phase are high (see text in
  Section 4 and Figure \ref{fig:implied_sfr}).}
\label{fig:age_mf_timescale}
\end{figure}

\begin{figure}
$\begin{array}{c}
\includegraphics[width=3.5in]{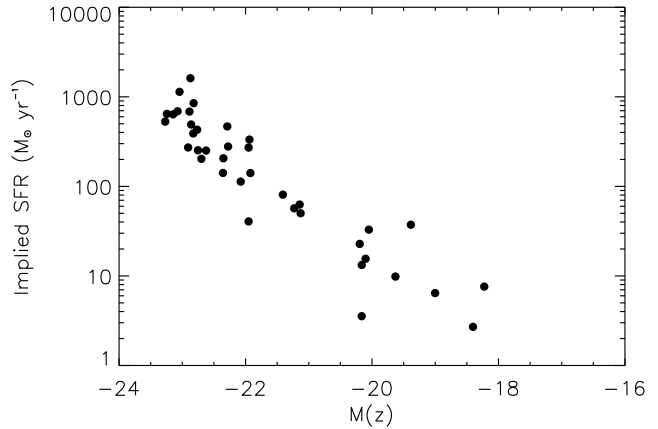}\\
\end{array}$
\caption{Implied SFRs during the burst phase in E+A galaxies. The
  implied SFR is estimated by dividing the stellar mass formed
  by the timescale of the burst. Note that,
  since the burst lasts longer than one timescale, the SFRs are
  overestimated. However, the estimates probably give a reasonable
  indication of the SFR during the most intense period of the
  starburst.}
\label{fig:implied_sfr}
\end{figure}


\section{Characteristics of the recent burst: ages, mass fractions and timescales}
We begin by presenting estimates for the basic parameters ($t_2$,
$f_2$ and $\tau_2$) that describe the burst of star formation that
immediately precedes the quenching in our E+A galaxies. The top
panel of Figure \ref{fig:age_mf_timescale} shows the ages and mass
fractions of the recent burst in each E+A galaxy. The burst ages
are typically within 2 Gyrs, as one expects from the lifetimes of
A-type stars which contribute significantly to E+A spectra. In
agreement with previous studies, we derive large mass fractions,
which are typically higher than 10 percent and possibly as high as
70 percent in some galaxies. The bottom panel shows the timescales
over which the bursts take place. The timescales are typically
short - between 0.01 and 0.2 Gyrs - implying that the star
formation rates during the burst phase are extremely high.

We can estimate this `implied' SFR by dividing the stellar mass
formed by the timescale of the burst (Figure
\ref{fig:implied_sfr}). We note that, since the burst lasts longer
than one timescale, the SFRs are overestimated. However, the
estimates probably give a reasonable indication of the SFR during
the most intense period of the starburst and we find a tight
correlation between the mass of the E+A galaxy and the implied
SFR. (see bottom panel of Figure \ref{fig:age_mf_timescale}).
While low-lumnosity ($M(z)>-20$) E+As have implied SFRs of less
than 50 $M_{\odot} yr^{-1}$, E+A systems at the high-luminosity
end ($M(z)<-22$) exhibit SFRs of greater than 300, and as high as
2000 $M_{\odot} yr^{-1}$.

The high SFRs inevitably lead to comparisons with luminous and
ultra-luminous infrared galaxies (LIRGs/ULIRGs), whose high
infrared luminosities ($L_{IR}>10^{11}L_{\odot}$) imply massive
ongoing starbursts \citep[e.g][]{Sanders1996}. SFRs in LIRGs
($10^{11}L_{\odot}<L_{IR}<10^{12}L_{\odot}$) typically exceed a
few tens of solar masses per year, while in ULIRGs
($L_{IR}>10^{12}L_{\odot}$) the SFRs can be as high as hundreds of
solar masses per year \citep[e.g.][]{Kennicutt1998}.
\citet{Wang2006}, who have performed a study of LIRGs in the
nearby Universe ($z\sim0.1$), find SFRs of between 10 and 100
$M_{\odot} yr^{-1}$ for their sample of galaxies (see their Figure
5). Noting that our SFRs are overestimated, and could be a few
factors too high, we suggest that LIRGs at low redshift
\emph{could} transform into massive E+A galaxies. However, we note
that the robustness of such a conclusion depends on performing an
identical parameter estimation on LIRGs at low redshift and by
considering whether other factors, such as the local environments
of E+A galaxies, correspond closely to those of LIRGs.


\begin{figure}
\begin{center}
$\begin{array}{c}
\includegraphics[width=3.5in]{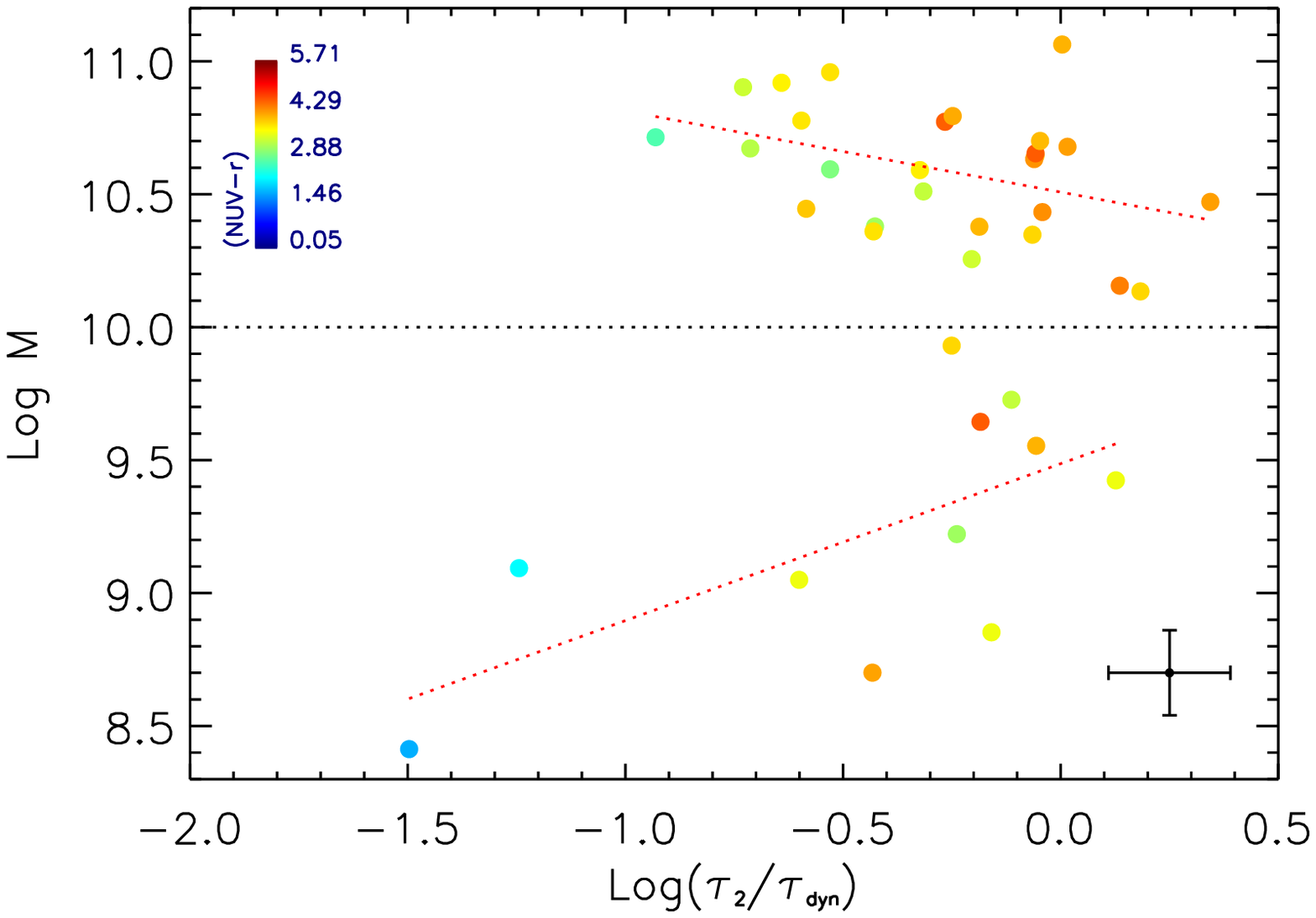}\\
\includegraphics[width=3.5in]{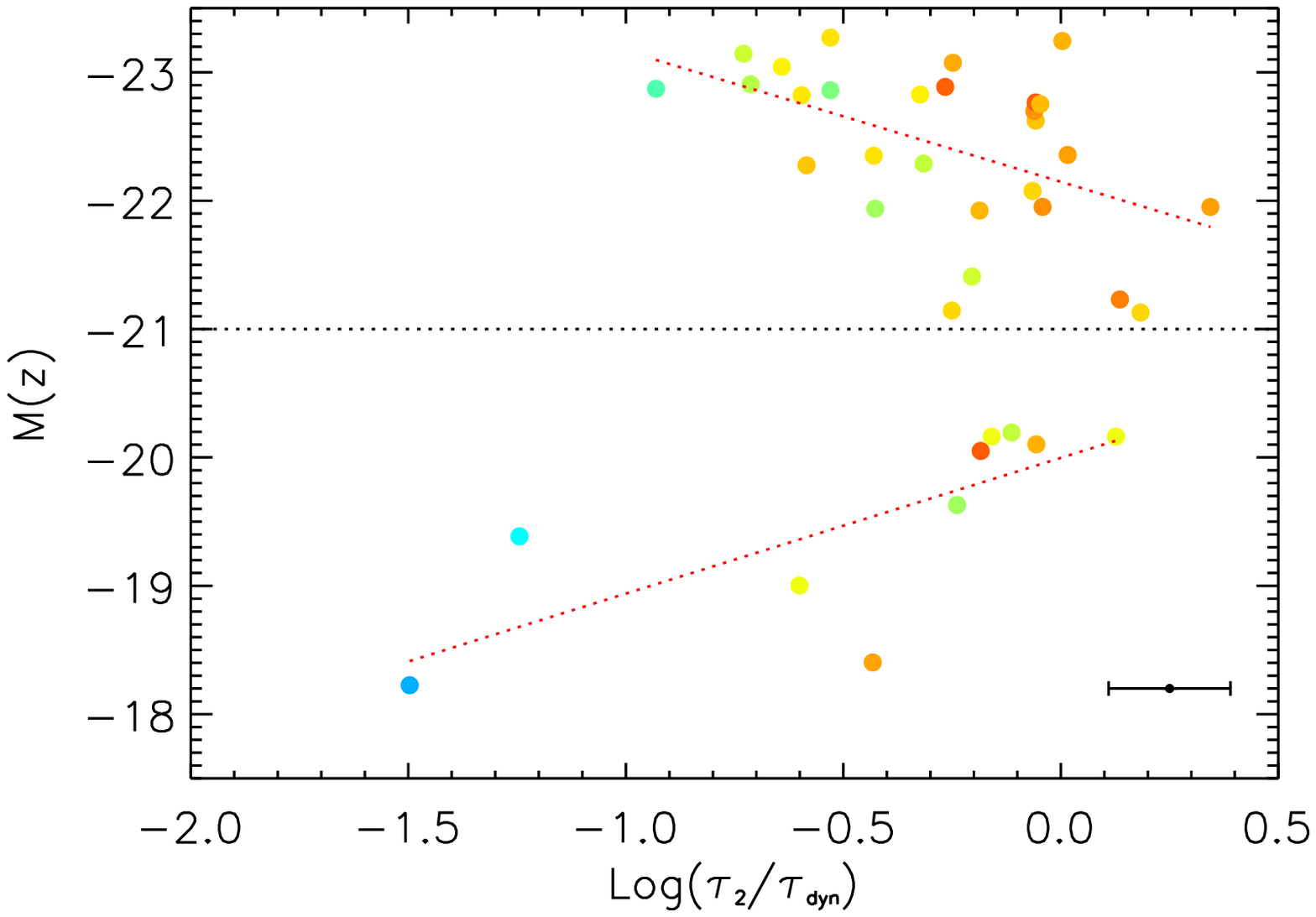}\\
\includegraphics[width=3.5in]{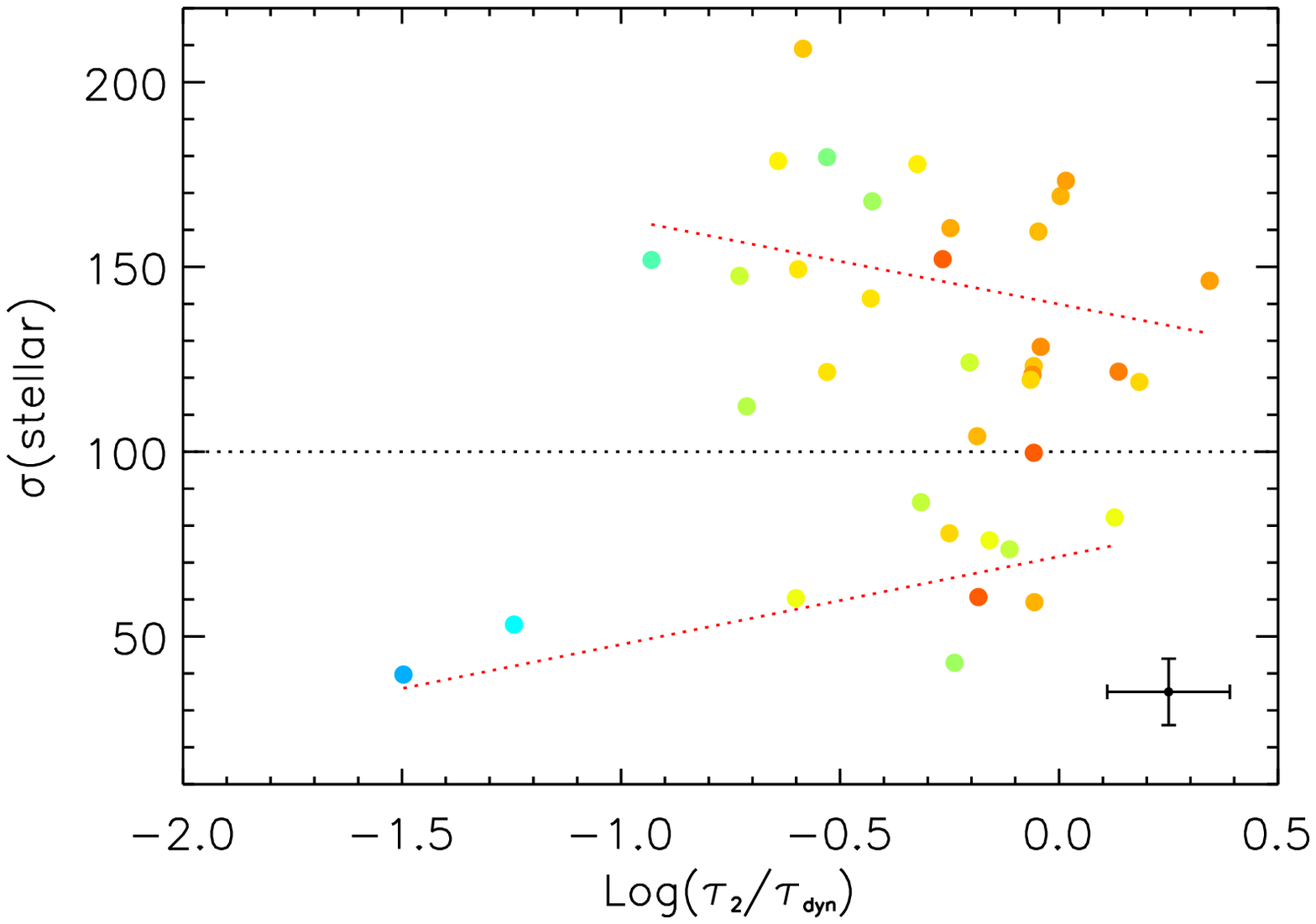}
\end{array}$
\caption{The timescale ratio, $r_{\tau} = \tau_2/\tau_{dyn}$,
plotted against galaxy mass (top), absolute z-band luminosity
(middle) and stellar velocity dispersion (bottom). Note that the
red dotted lines indicate standard least-squares fits to the data.
The fits are performed separately for the massive
($M>10^{10}M_{\odot}$) and low-mass ($M<10^{10}M_{\odot}$) E+A
galaxies in each panel.}\label{fig:timescale_ratio}
\end{center}
\end{figure}


\section{The efficiency of quenching as a function of galaxy mass}
One of the primary goals of this paper is to study the
characteristics of the quenching that truncates the burst and
determine the possible mechanisms by which this quenching could
occur. A key parameter in this analysis is the \emph{efficiency}
with which the star formation is quenched. We study this
`quenching efficiency' by comparing the effective timescale of the
burst, $\tau_2$, to the dynamical timescale\footnote{Using
$s=\frac{1}{2}at^2$ from the equations of motion, the dynamical
timescale follows by setting $a=g=(GM/R^2)$ and $s=R$.} of the
galaxy, $\tau_{dyn}$, which is defined as

\begin{equation}
\tau_{dyn} = \Big(\frac{2R^3}{GM}\Big)^{1/2},
\end{equation}

and describes the `natural' timescale over which processes such as
star formation would take place if left \emph{unhindered}
\footnote{Note that we use the radius that contains 90 percent of the
  Petrosian flux in the $i$-band as the value of $R$.}. Note that, in the calculations that follow, we have used the stellar mass of the galaxy as the value of $M$. We
parametrise the quenching efficiency by constructing the timescale
ratio

\begin{equation}
r_{\tau} = \tau_2/\tau_{dyn}.
\end{equation}

More efficient quenching reduces $\tau_2$ and therefore produces a
lower value for $r_{\tau}$. The timescale ratio is, therefore,
\emph{inversely correlated} with quenching efficiency - more
catastrophic quenching results in lower values of $r_{\tau}$.

In Figure \ref{fig:timescale_ratio} we plot $r_{\tau}$ against
three independent indicators of galaxy mass/luminosity. The
galaxies are colour-coded using their $(NUV-r)$ colour. The plot
of $r_{\tau}$ vs galaxy mass (top panel), reveals a striking
difference in the behaviour of $r_{\tau}$ with increasing galaxy
mass. We find that, below a mass of $10^{10}M_{\odot}$, the
quenching efficiency decreases with increasing galaxy mass.
However, when the galaxy mass is greater than
$\sim10^{10}M_{\odot}$ this trend is reversed and the quenching
efficiency then correlates positively with the galaxy mass. We
check this behaviour with both $M(z)$ and stellar velocity
dispersion. We find that the correlation of $r_{\tau}$ with galaxy
mass and, in particular, its apparent reversal is present in all
the separate indicators of galaxy mass/luminosity. While a strong
trend is apparent with stellar mass ($M$) and $z$ band luminosity,
the trend is weaker in the stellar velocity dispersion ($\sigma$),
most probably driven by bigger uncertainties in the $\sigma$
measurement (as suggested by the larger scatter in this plot). A
plausible source of uncertainty is the fact that the E+A galaxies
are unrelaxed (indeed a significant fraction look disturbed in the
SDSS images), implying that $\sigma$ does not yet trace the virial
mass of the galaxy. An additional source of uncertainty is the
presence of a substantial young stellar population. The presence
of young sub-components could create a bias towards lower values
of $\sigma$, with the magnitude of the bias driven by the mass
fraction and age of the young stars and on the exact size of the
physical aperture subtended by the SDSS spectra in each galaxy. We
therefore suggest that the correlation between $r_{\tau}$ and
$\sigma$ should be treated with caution considering the caveats we
have presented above.

Figure \ref{fig:timescale_ratio} shows that the reversal in the
$r_{\tau}$ correlation occurs at $M\sim10^{10}M_{\odot}$,
$M(z)\sim-21$ and $\sigma\sim100$ kms$^{-1}$ when $r_{\tau}$ is
plotted against mass, absolute z-band luminosity and stellar
velocity dispersion respectively. The dichotomy in the trend of
$r_{\tau}$ against galaxy mass indicates that the principal
quenching mechanisms for galaxies above and below the mass
threshold of $10^{10}M_{\odot}$ behave very differently. In
particular, the mechanism that operates in the regime
$M<10^{10}M_{\odot}$ becomes \emph{weaker} as the galaxy mass
increases while the mechanism that operates in the regime
$M>10^{10}M_{\odot}$ becomes \emph{stronger} as the galaxy mass
increases.

The mass threshold ($10^{10}M_{\odot}$) at which the $r_{\tau}$
vs. $M$ trend reverses is particularly interesting because it is
in excellent agreement with the mass above which AGN become
significantly more abundant in galaxies in the nearby Universe
(Kauffmann et al. 2003a). Given this result, the reversal at
$M\sim10^{10}M_{\odot}$ can be explained, at least qualitatively,
in terms of SN and AGN being the principal quenching sources above
and below this mass threshold. In the absence of AGN, the primary
source of negative feedback are SN. As galaxies become more
massive and the depth of the potential well increases, SN find it
increasingly more difficult to eject gas from the system. This
results in the quenching becoming less efficient at higher galaxy
masses, exactly as we observe in Figure \ref{fig:timescale_ratio}.

However, once AGN begin to appear (above $M\sim10^{10}M_{\odot}$),
they become the dominant source of negative feedback. For a
$10^{11}M_{\odot}$ galaxy, the energy input from a population of
(Type II) SN formed as a result of a starburst that creates half
the stellar mass of the galaxy is $\sim 3 \times 10^{57}$
ergs\footnote{The energy injected by a population of SN is given
by $\eta_{SN}\varepsilon_{SN}\Delta M$, where $\eta_{SN} (\sim
5\times 10^{-3}M_{\odot}^{-1})$ is the number of SN per unit
stellar mass and $\varepsilon_{SN} (\sim 10^{51}$ ergs) is the
energy supplied by each supernova. The fraction of that energy
that couples mechanically to the material in the ISM is
approximately given by $\sigma/20,000$ $kms^{-1}$, where $\sigma$
is the velocity dispersion and 20,000 $kms^{-1}$ is the velocity
of the SN ejecta. For a $M\sim10^{11}M_{\odot}$ galaxy,
$\sigma\sim200$ $kms^{-1}$.}, while the energy output from a
central AGN, emitting at the Eddington limit\footnote{The
Eddington luminosity for a mass $M$ is $\sim
1.3\times10^{38}M/M_{\odot}$ ergs per second.} over a timescale of
0.1 Gyrs is $\sim 3 \times 10^{58}$ ergs\footnote{A
$M\sim10^{11}M_{\odot}$ typically hosts a $M\sim10^{8}M_{\odot}$
black hole, since $(M_{BH}/M_{bulge}) \sim 10^{-3}$
\citep{Haring2004}. The Eddington luminosity of such a black hole
is $\sim 10^{46} ergs$
 $s^{-1}$, of which a fraction $\sigma/c$ $kms^{-1}$ couples to the
mechanically to the material in the ISM. $\sigma$ is the velocity
dispersion of the galaxy and $c$ is the speed of light.}. In terms
of energy input the AGN dominates and it is reasonable to assume
that it dictates the mechanics of the quenching. Note that we did
not consider Type Ia supernovae in this calculation because the
derived timescales (see Figure \ref{fig:age_mf_timescale}) are
considerably shorter than 1 Gyr, which is the typical time delay
between the onset of star formation and the appearance of Type Ia
supernovae.

Given that the mass of the black hole ($M_{BH}$) scales with the
central velocity dispersion $\sigma$ as $M_{BH}\sim\sigma^{\beta}$
\citep{Gebhardt2000,Ferrarese2000}, where $\beta$ varies in range
4-5, and assuming the energy outputted by the AGN scales with
$M_{BH}$ (c.f. the Eddington luminosity $L_{edd}\propto M_{BH}$),
we expect AGN feedback to become more effective, and thus the
quenching efficiency to increase, as galaxy mass increases. Again,
this is indeed what we observe in Figure
\ref{fig:timescale_ratio}. We briefly note that, by definition,
our sample does not contain galaxies that show signs of
\emph{current} AGN activity, implying that the AGN has shut itself
off in the process of providing feedback. If, as we argue below,
mechanical feedback from the AGN depletes the gas reservoir, then
this process also removes the AGN's own fuel source. Thus it is
plausible that the feedback process \emph{simultaneously} quenches
both star formation and AGN activity.

It is worth noting that a similar dichotomy in the primary
feedback mechanism has been inferred by \citep{Shankar2006}, from
a study of the stellar and baryonic mass functions of galaxies,
extracted using the mass-to-light ratios of stars and gas derived
from galaxy kinematics.


\section{The expected dependence of the quenching efficiency on galaxy mass}
We now use simple energetic arguments to derive the expected
relationship between the timescale ratio $r_{\tau}$ and galaxy mass in
the two cases where the quenching is due to SN and AGN respectively.

We make the assumption that quenching takes place due to the
ejection of available gas from the potential well. The energy
required for quenching, $E_q$, is given by,

\begin{equation}
E_q=\frac{1}{2}m_gv_{esc}^2,
\end{equation}

where $m_g$ is the mass of gas to be ejected and $v_{esc}^2$ is
the escape velocity. Since,

\begin{equation}
v_{esc}^2=\Big(\frac{2GM}{R}\Big),
\end{equation}

this implies that

\begin{equation}
E_q\propto m_g.\frac{M}{R},
\end{equation}

or equivalently, in terms of parameters derived in this analysis
that,

\begin{equation}
E_q \propto m_g.\Big(\frac{M}{\tau_{dyn}}\Big)^{2/3} \propto
m_g.\Big(\frac{Mr_{\tau}}{\tau_2}\Big)^{2/3}.
\end{equation}

The relationships in Eqn (6) use substitutions from Eqns (1) and
(2). Eqn (6) describes the dependence of the energy required for
quenching on the timescale ratio $r_{\tau}$, the burst timescale
$\tau_2$ and galaxy mass $M$.

We now make the simplifying assumption that the mass of gas
($m_g$) required to be ejected for quenching is approximately
constant in all galaxies. This is consistent with the downsizing
phenomenon \citep[e.g][]{Cowie1996}, whereby smaller galaxies
exhibit higher star formation rates than their massive
counterparts at lower redshifts. Given the apparent universality
of the Schmidt-Kennicutt law for (quiescent) star formation
\citep{Kennicutt1998}, downsizing implies that smaller galaxies
tend to be more gas-rich than larger ones, justifying the
empirical assumption that their \emph{absolute} neutral gas
reservoirs may have similar sizes. Cold gas fractions in
low-redshift LIRGs \citep[see Figure 5 in][]{Wang2006}, which are
potential progenitors of E+A galaxies (Section 4), indeed show a
declining trend with galaxy mass. Across the mass range
$10^{10.5}-10^{11.5}M_{\odot}$, the gas fractions increase from
less than 5 percent to greater than 25 percent, suggesting that
the cold gas reservoirs in these galaxies are similar in size and
that our assumption of constant $m_g$ is a reasonable one. A small
spread in the real values of $m_g$ will induce a scatter in the
trends derived in Section 6.1 and 6.2 below, without perturbing
the correlations between the variables. With this assumption Eqn
(6) becomes

\begin{equation}
E_q \propto \Big(\frac{M}{\tau_{dyn}}\Big)^{2/3} \propto
\Big(\frac{Mr_{\tau}}{\tau_2}\Big)^{2/3}.
\end{equation}

We should also note that the assumption that quenching occurs
through the mechanical ejection of gas may imply that E+A galaxies
should be relatively gas-poor, although the gas could simply be
dispersed away from the central regions of the potential well
where star formation would normally take place. Two studies of the
$HI$ content of E+A galaxies have been conducted so far.
\citet{Chang2001} observed five E+A galaxies from the
\citet{Zabludoff1996} sample and detected only one with an $HI$
mass of $\sim3.4\times10^9M_{\odot}$. Similarly, \citet{Buyle2006}
detected similar amounts ($>10^9M_{\odot}$) of $HI$ in three out
of six E+A galaxies they observed in $HI$. While, it might be
useful to compare the gas masses estimated by
\citet[e.g][]{Buyle2006} for E+A galaxies to those estimated for
LIRGs (as potential progenitors) by \citet{Wang2006}, the mass
ranges traced by these two studies are quite different.
\citet{Buyle2006} trace galaxies with
$M\lesssim10^{10.7}M_{\odot}$\footnote{The masses of the
  \citet{Buyle2006} galaxies have been estimated by converting the
  values of $M(B)$ given in their Table 1 to $M(z)$, which are then
  converted to masses using the $M$ vs. $M(z)$ relation for the E+A galaxies
  in our sample.}, while \citet{Wang2006} study galaxies with
$M\gtrsim10^{10.6}M_{\odot}$). The small overlap region around
$10^{10.7}M_{\odot}$ indicates that the $HI$
fraction\footnote{Note that
  we have converted the $H_2$ mass (which is the quantity estimated by \citet{Wang2006}) to an $HI$ mass, using the average $HI/H_2$ ratio for late-type
  galaxies ($\sim0.6$) given by \citet[][see their Table 1]{Fukugita1998}} in the LIRGs may be higher by a factor of 3.
  We note, however, that due to the inadequate overlap between the two
samples, this result is not very robust. However, given the
apparent lack of $HI$ in the majority of E+A's observed by
\citet{Chang2001} and the smaller gas fractions found in
\citet{Buyle2006} compared to nearby LIRGs (which have comparable
SFRs to our E+A galaxies) the assumption of quenching through the
mechanical ejection of gas seems reasonable.


\subsection{Quenching by supernovae}

\begin{figure}
\begin{center}
\includegraphics[width=3.5in]{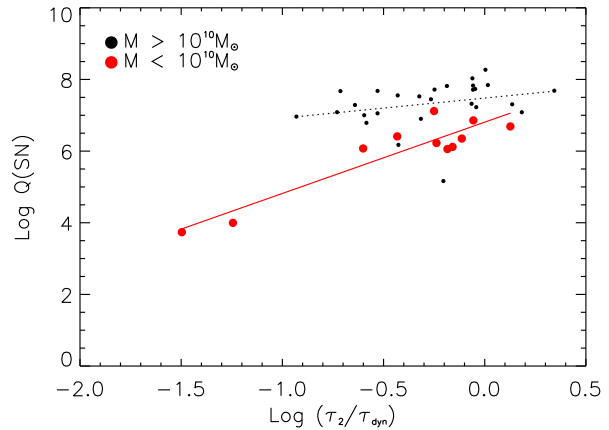}
\caption{$r_{\tau}$ vs. $Q_{SN}$ for galaxies with
$M>10^{10}M_{\odot}$ (black) and $M<10^{10}M_{\odot}$ (red). The
solid lines indicate standard least-squares fits to the data.}
\label{fig:snfeedback}
\end{center}
\end{figure}

The energy injected by SN into the ISM is proportional to the
amount of stellar mass produced in the burst so that

\begin{equation}
E^{SN}=\eta_{SN}\varepsilon_{SN}\Delta M,
\end{equation}

where $\eta_{SN}$ is the number of SN per unit stellar mass,
$\varepsilon_{SN}$ is the energy supplied by each supernova and
$\Delta M$ is the amount of stellar mass formed. Hence,

\begin{equation}
E^{SN} \propto f_2M,
\end{equation}

since $\Delta M=f_2M$. Combining Eqn (7) and Eqn (9) gives,

\begin{equation}
r_{\tau}^2\propto\tau_2^2f_2^3M,
\end{equation}

so that

\begin{equation}
\log(\tau_2^2f_2^3M) = 2\log(r_{\tau}) + k
\end{equation}

As we would naively expect from SN driven feedback, $r_{\tau}$
correlates with the galaxy mass ($M$). Since $r_{\tau}$ is
inversely correlated to the quenching efficiency, this implies
that, as galaxy mass increases, the quenching efficiency decreases
in the SN feedback scenario. Defining $Q_{SN}=\tau_2^2f_2^3M$ we
have,

\begin{equation}
\log(Q_{SN}) = 2\log(r_{\tau}) + k,
\end{equation}

where $k$ absorbs the constants in the proportionality relations
used to derive Eqn (12). A plot of $\log(Q_{SN})$ against
$\log(r_{\tau})$ should therefore show a gradient of 2, if the
systems being considered are indeed quenched \emph{purely} by SN.

In Figure \ref{fig:snfeedback} we plot $r_{\tau}$ vs. $Q_{SN}$ for
both low-mass ($M<10^{10}M_{\odot}$; black) and massive
($M>10^{10}M_{\odot}$; red) galaxies. The solid lines indicate
standard least-squares fits to the data. We find that, while
low-mass galaxies show a gradient of $1.98\pm0.18$ (and are
therefore consistent with the expected gradient of 2), the
best-fit relation for the massive galaxies is too shallow
($0.56\pm0.06$) and does not satisfy the predictions. We conclude,
therefore, that low-mass galaxies show good consistency with being
quenched \emph{purely} by SN feedback while massive galaxies are
essentially inconsistent with such a picture.

\begin{figure}
\begin{center}
\includegraphics[width=3.5in]{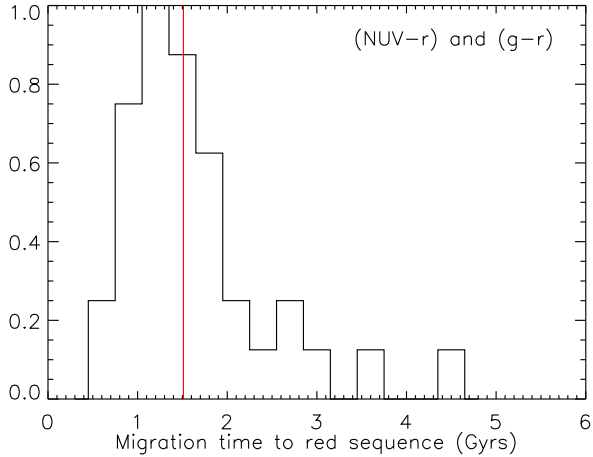}\\
\caption{Histogram showing the migration time of E+A galaxies from
the blue cloud to the red sequence in both the $UV$ and optical
colours. The median migration time (indicated by the solid red
line is $\sim1.5$ Gyrs. The y-axis shows the frequency normalised
to 1.)} \label{fig:nr_gr_migration}
\end{center}
\end{figure}


\subsection{Quenching by AGN} The energy injected by an AGN
($E^{AGN}$) is driven by its luminosity ($L$) so that,

\begin{equation}
E^{AGN} \propto L.\tau_2.
\end{equation}

Rewriting Eqn (13) in terms of the Eddington luminosity
($L_{edd}$), we have

\begin{equation}
E^{AGN} \propto \mu.L_{edd}.\tau_2,
\end{equation}

where $\mu=L/L_{edd}$ is the Eddington ratio. $L_{edd}$ is
proportional to the mass of the central black hole ($M_{BH}$),
which in turn has been shown to scale with the central velocity
dispersion ($\sigma$) of the galaxy as $M_{BH} = \sigma^{\beta}$.
Estimates of $\beta$ vary in the range 4-5, with
\citet[e.g.][]{Gebhardt2000} reporting $\beta=3.75 \pm 0.3$, while
\citet{Ferrarese2000} report $\beta=4.8 \pm 0.5$. For simplicity,
we begin by assuming that $\mu$ is constant across our entire
sample of AGN. Then,

\begin{equation}
E^{AGN} \propto \sigma^{\beta}.\tau_2.
\end{equation}

Since, the virial theorem\footnote{The virial theorem states that,
for
  a system with kinetic energy $K$ and potential energy $U$,
  $2K+U=0$. For a mass distribution of mass $M$ and radius $R$,
  $K\sim\frac{1}{2}M\sigma^2$ and $U\sim\frac{GM}{R^2}$ which implies
  that $\sigma^2 \sim M/R$.} implies that $\sigma^2 \sim M/R$ we
rewrite this as,

\begin{equation}
E^{AGN} \propto \frac{M^{\beta/3}r_{\tau}^{\beta/3}}{\tau_2^{\beta/3}}.\tau_2,
\end{equation}

where we have used Eqns (1) and (2) to eliminate $R$ and
explicitly introduce the timescale ratio $r_{\tau}$. Equating Eqn
(7) and Eqn (16) then gives

\begin{equation}
r_{\tau}^{(2-\beta)} \propto M^{(\beta-2)}.\tau_2^{(5-\beta)}
\end{equation}

Using $\beta \sim 5$ from \citet{Ferrarese2000} then implies that

\begin{equation}
M  \propto r_{\tau}^{-1}
\end{equation}

or that

\begin{equation}
\log{M}= -\log(r_{\tau})+k,
\end{equation}

where $k$ absorbs the constants in the proportionality relations
used to derive Eqn (18). We find that a plot of $\log{M}$ against
$\log{r_{\tau}}$ should show a gradient of -1, if the systems
being considered are indeed quenched by AGN which have similar
Eddington ratios, regardless of their size. Eqn (18) explicitly
predicts that, for galaxies which are quenched by AGN alone, the
quenching efficiency (which is inversely correlated to $r_{\tau}$)
increases as the mass of the galaxy increases.

We now refer back to Figure \ref{fig:timescale_ratio} (top panel),
where we have plotted the dependence of $r_{\tau}$ on galaxy mass
for both massive and low-mass galaxies. It is clear that low-mass
galaxies are inconsistent with the energetics of AGN-driven
quenching, since for these systems $r_{\tau}$ correlates
positively with galaxy mass. Massive galaxies, on the other hand,
do show an inverse correlation with $r_{\tau}$. However, the
observed gradient in the correlation in Figure
\ref{fig:timescale_ratio} is $-0.36 \pm 0.11$, shallower than the
predicted gradient of -1.

This discrepancy arises due to our assumption, in Eqn (13), that
the Eddington ratio during the burst phase ($\mu$) does not change
as a function of host galaxy mass. Clearly, if $\mu$ decreases,
then the rate of energy output ($L$) also decreases and a longer
timescale ($\tau_2$) is required to inject the same amount of
energy. This, in turn, implies a larger value of $r_\tau$.

We now refine our previous analysis by allowing the Eddington
ratio to vary as a function of mass so that,

\begin{equation}
\mu=M^{\gamma}
\end{equation}

With this dependence on $M$, Eqn (16) becomes

\begin{equation}
E^{AGN} \propto
\frac{M^{\beta/3}r_{\tau}^{\beta/3}}{\tau_2^{\beta/3}}.\tau_2.M^{\gamma},
\end{equation}

so that,

\begin{equation}
r_{\tau}^{(2-\beta)} \propto
M^{(3{\gamma}+\beta-2)}.\tau_2^{(5-\beta)}.
\end{equation}

Using $\beta \sim 5$ as before then gives,

\begin{equation}
M \propto r_{\tau}^{-1/(1+\gamma)}.
\end{equation}

Comparison with the observed gradient of $-0.36 \pm 0.11$ implies
that $1<\gamma<3$. We find, therefore, that the observed gradient
in the $\log M$ vs. $\log{r_{\tau}}$ correlation can be reproduced
if the Eddington ratio of the E+A galaxies is assumed to vary with
$M$. Since the E+A masses are spread over an order of magnitude,
this implies that the Eddington ratios of the smallest AGN (with
host galaxy masses $M\sim10^{10}M_{\odot}$) in this sample of
galaxies are \emph{at least} one-tenth of those for their most
massive counterparts in this sample (if $\gamma\sim1$).

Finally, it is instructive to calculate if a variable Eddington
ratio (with $1<\gamma<3$) might indeed be expected from simple
arguments. The luminosity ($L$) of an AGN is proportional to the
mass accretion rate (\emph{\.{M}}). Spherical, Bondi-Hoyle
accretion \citep{Bondi1944} implies that \emph{\.{M}} is
proportional to the square of the mass of the central black-hole
(BH) i.e.

\begin{equation}
L \propto \textnormal{\emph{\.{M}}} \propto M_{BH}^2.
\end{equation}

Since the Eddington luminosity ($L_{edd}$) is proportional to
$M_{BH}$ we have,

\begin{equation}
\frac{L}{L_{edd}} \propto M_{BH}.
\end{equation}

The BH mass is typically a constant fraction of the mass of the
bulge \citep[e.g.][]{Haring2004}, with

\begin{equation}
\frac{M_{BH}}{M_{bulge}} \sim 10^{-3},
\end{equation}

which implies that,

\begin{equation}
\mu = \frac{L}{L_{edd}} \propto M,
\end{equation}

where we have assumed that the E+A galaxies in our sample are
bulge-dominated. Such a scenario implies that the expected value
of $\gamma$ in the analysis presented above is $\sim1$, which is
consistent with the derived range ($1<\gamma<3$).

Assuming a more general dependence of \emph{\.{M}} on $M_{BH}$
($\textnormal{\emph{\.{M}}} \propto M_{BH}^\lambda$), would imply
that,

\begin{equation}
\frac{L}{L_{edd}} \propto M^{\lambda-1}.
\end{equation}

$\lambda-1$ is equal to $\gamma$ in Eqn (20) and the Eddington
ratio increases with host galaxy mass (as is required to satisfy
the shallower gradient in the $\log M$ vs. $\log r_{\tau}$
relation) if $\lambda\gtrsim 2$. For Bondi-Hoyle accretion
$\lambda\sim2$.



\section{Migration time to the red sequence}
Galaxy colours show a pronounced bimodality over a large range in
redshift. This dichotomy in the colours, combined with the
build-up of the red sequence over time \citep[e.g.][]{Bell2004},
indicates that a central feature of the evolution of the galaxy
population is the net migration of galaxies from the blue cloud
onto the red sequence. Therefore, in the context of understanding
the macroscopic evolution of galaxies, it is useful to have an
estimate of the typical time it takes to complete this
migration.

\begin{figure}
\begin{center}
$\begin{array}{c}
\includegraphics[width=3.5in]{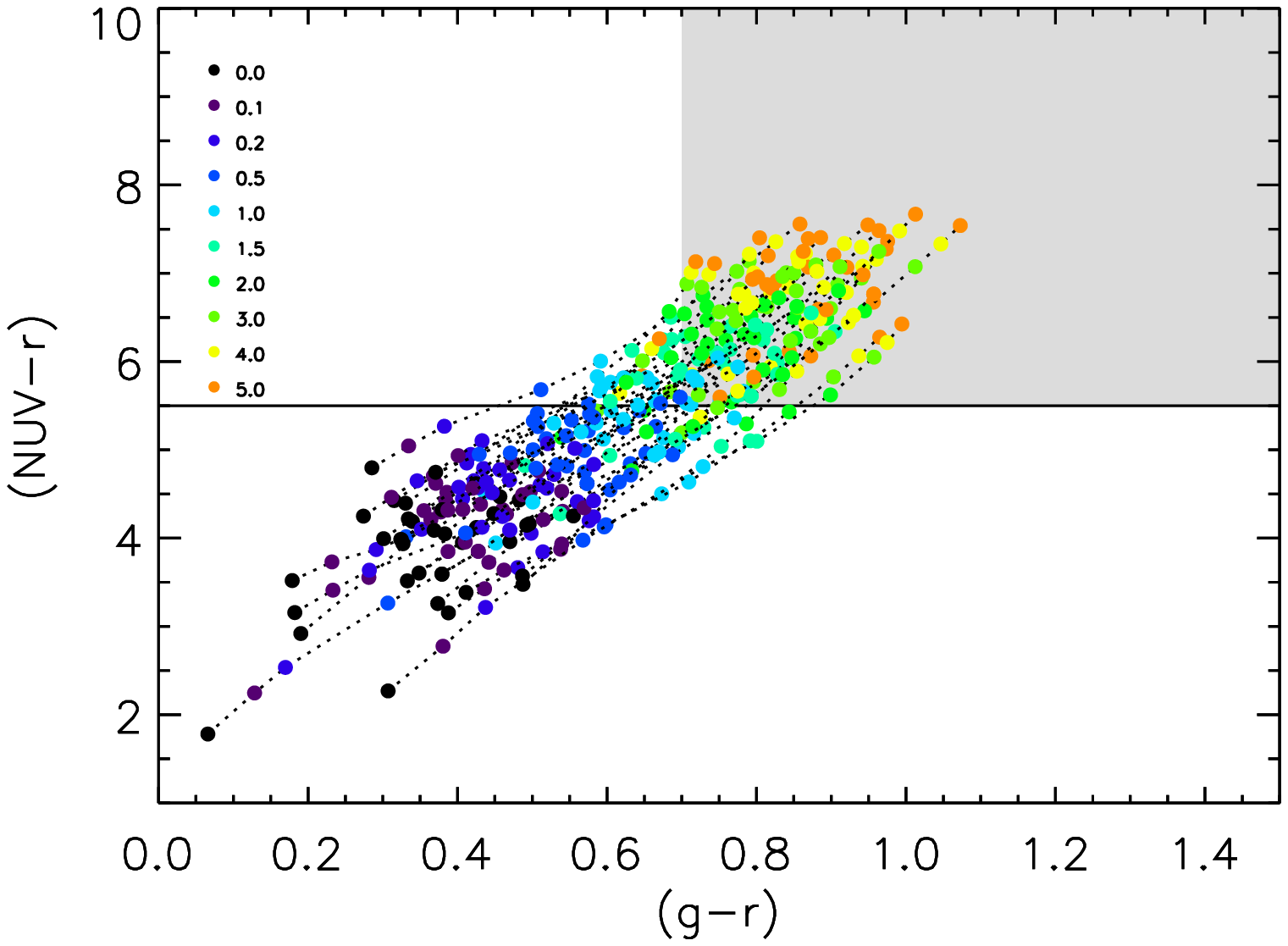}\\
\includegraphics[width=3.5in]{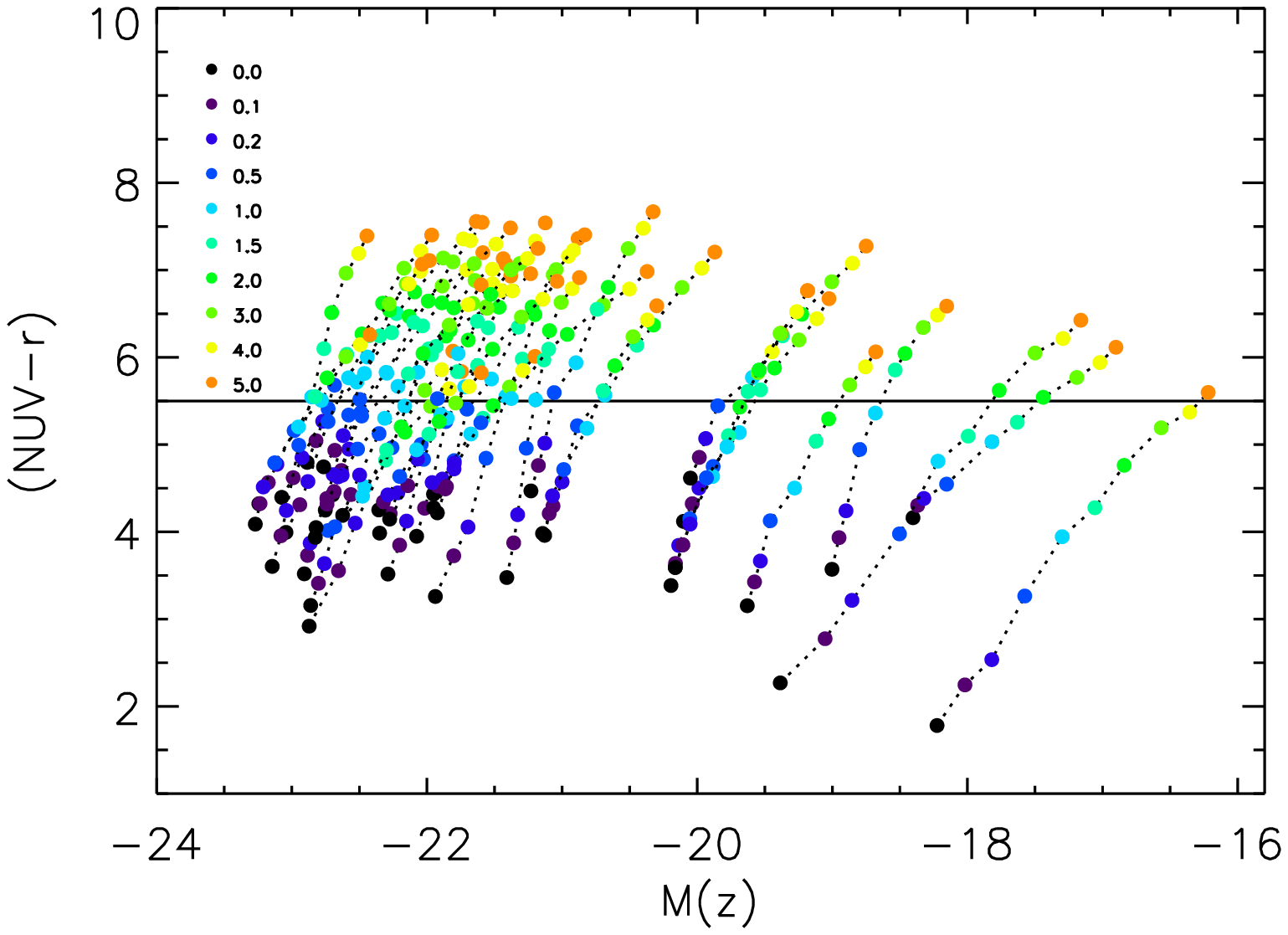}\\
\end{array}$
\end{center}
\caption{Migration tracks of E+A galaxies in the $(NUV-r)$ vs
$(g-r)$ colour space (top panel) and the $(NUV-r)$ vs $M(z)$
colour-magnitude space (bottom panel). Ages (in Gyrs) along the
track are shown colour-coded.} \label{fig:migration_tracks}
\end{figure}

Given the very recent truncation of star formation in these
systems, E+A galaxies are just about to begin this migration
process. Assuming there is negligible star formation during this
phase, we can use our derived SFHs to estimate the time it takes
for galaxies to move from the blue cloud to the red sequence in
\emph{both} the $UV$ and optical colours. The use of E+A galaxies
and the fact that we study the migration in \emph{both} the $UV$
and optical spectrum makes the derived migration times quite
robust. Note that we define the red sequence as the part of the
$(NUV-r)$ vs $(g-r)$ colour space that hosts the bulk of the
\emph{red} early-type population. Comparison to Figure
\ref{fig:cmspace} indicates that the red sequence lies redward of
$(NUV-r)\sim5.5$ and $(g-r)\sim0.7$.

We estimate the migration times by `ageing' the \emph{best-fit}
model SFHs of each E+A galaxy, keeping secondary parameters such
as metallicity and $E(B-V)$ constant. We find that the migration
times vary between 1 and 5 Gyrs (Figure
\ref{fig:nr_gr_migration}). Most galaxies complete their migration
within 2 Gyrs and the median migration time of galaxies in this
sample (shown using the solid red line) is $\sim1.5$ Gyrs.
Finally, in Figure \ref{fig:migration_tracks} we show the
migration tracks of the E+A sample in the $(NUV-r)$ vs $(g-r)$
colour space (top panel) and the $(NUV-r)$ vs $M(z)$
colour-magnitude space (bottom panel). Ages along the track are
shown colour-coded.


\section{Conclusions}
We have presented the first study of nearby E+A galaxies which
incorporates their $UV$ photometry. By exploiting the sensitivity
of the $UV$ to young stars, we have accurately reconstructed the
recent star formation histories of 38 E+A galaxies in the nearby
Universe ($0<z<0.2$), by combing optical ($u,g,r,i,z$) and $UV$
data from the SDSS and GALEX surveys respectively.

We find that the burst of star formation that dominates the
post-starburst signatures in these galaxies typically takes place
within a Gyr, which is consistent with the presence of A-type
stars in these systems. The stellar mass fractions formed in this
burst are typically high, ranging from 20 percent to 60 percent of
the mass of the E+A remnant. The timescale over which this star
formation takes place is short, ranging between 0.01 and 0.2 Gyrs.
The combination of short timescales and high mass fractions imply
high SFRs during the burst. We find a tight, positive correlation
between the mass of the E+A galaxy and the implied SFR. While
low-luminosity ($M(z)>-20$) E+As have implied SFRs of less than 50
$M_{\odot} yr^{-1}$, E+A systems at the high-luminosity end
($M(z)<-22$) exhibit SFRs of greater than 300, and as high as 2000
$M_{\odot} yr^{-1}$. The SFRs are comparable to those found in
LIRGs and ULIRGs at low redshift and our results indicate that
massive LIRGs at low redshift could be the progenitors of massive
E+A galaxies like those found in our sample.

We have performed a comprehensive study of the characteristics of
the quenching that truncates the starburst in E+A galaxies. In
particular, we have studied how the quenching efficiency varies as a
function of galaxy mass and compared the results to
scenarios where the quenching is due to SN and AGN, which are the
typical sources of kinetic and thermal feedback invoked in galaxy
formation models. We have found that in E+A galaxies with masses
lesser than $10^{10}M_{\odot}$, quenching becomes less efficient
as the galaxy mass increases. However, in galaxies with masses
greater than $10^{10}M_{\odot}$, this trend is reversed and the
quenching efficiency scales positively with galaxy mass. In terms of
$M(z)$ and stellar velocity dispersion ($\sigma$), the reversal occurs
at $M(z)\sim-21$ and $\sigma\sim100$ kms$^{-1}$.

Since the mass threshold ($10^{10}M_{\odot}$) where this reversal
occurs is in excellent agreement with the mass above which AGN
become significantly more abundant in nearby galaxies, these
results can be qualitatively explained in terms of AGN and SN
being the principal sources of feedback that quenches star
formation in E+A galaxies above and below $M\sim10^{10}M_{\odot}$
respectively. In the absence of AGN ($M<10^{10}M_{\odot}$), the
primary source of negative feedback are SN. As galaxies become
more massive, the increasing depth of the potential well makes it
more difficult to eject gas from the system, reducing the
quenching efficiency. As a result, the quenching efficiency shows
a negative correlation with galaxy mass. We have shown that simple
energetic arguments, based on the assumption that quenching occurs
through mechanical ejection of gas, are able to satisfy the
observed properties of E+A galaxies with masses below
$10^{10}M_{\odot}$. This indicates that, in these galaxies, the
quenching is likely to be driven \emph{purely} through mechanical
feedback from SN.

Once they begin to appear ($M>10^{10}M_{\odot}$), AGN become the
dominant source of energetic feedback. Given that the AGN
luminosity scales with the mass of the black hole ($M_{BH}$)
which, in turn, scales strongly with the central velocity
dispersion $\sigma$ as $M_{BH}\sim\sigma^{4-5}$, we expect AGN
feedback to become more effective (and hence the quenching
efficiency to increase) as the galaxy mass increases. We have
shown that simple energetic arguments, based on the mechanical
ejection of gas, indeed expect a positive correlation between the
quenching efficiency and galaxy mass. However, the derived
properties suggest that the quenching efficiencies rise an order
of magnitude faster with galaxy mass than predicted by simple
energetic arguments \emph{that assume that the Eddington ratio
does not vary as a function of host galaxy mass}. However, the
correlation between quenching efficiency and galaxy mass can be
reproduced if the Eddington ratio ($\mu$) of the E+A galaxies is
assumed to vary with $M$ as $M^{\gamma}$, where $1<\gamma<3$.
Since the E+A masses are spread over an order of magnitude, this
implies that the Eddington ratios of the smallest AGN (with host
galaxy masses $M\sim10^{10}M_{\odot}$) are roughly one-tenth of
those for their most massive counterparts in this sample (if
$\gamma\sim1$).

Finally, we have used our E+A sample to estimate the time it takes
for galaxies to migrate from the blue cloud to red sequence. The
persistent bimodality in galaxy colours over a large range in
redshift coupled with the build-up of the red sequence over time
suggests that the net migration of galaxies from the blue cloud to
the red sequence is an important feature of the macroscopic
evolution of the galaxy population over time. The use of E+A
galaxies, which have just truncated their star formation and are
on the verge of this migrationary transition, together with the
use of both $UV$ and optical photometry, produces reasonably
robust estimates of the migration times. The migration times are
estimated by `ageing' the best-fit SEDs of each E+A galaxy,
keeping secondary parameters such as metallicity and $E(B-V)$
constant throughout the migration. We calculate migration times
between 1 and 5 Gyrs, with a typical migration time of $\sim1.5$
Gyrs.

The study of E+A galaxies arguably holds the key to understanding
many of the processes that shape the evolution of galaxies. While
our analysis is phenomenological in nature, this study has
attempted to derive quantitative insights into some of these
processes, especially the characteristics of feedback that is
responsible for modulating and truncating star formation in
galaxies. While many of the results derived here (e.g. timescales
over which AGN or SN feedback quenches star formation, the
comparative efficiencies of the two feedback modes in potential
wells of varying sizes or the migration times from the blue cloud
to the red sequence) could prove useful constraints in galaxy
formation models, future studies of E+A galaxies at high redshift
are keenly anticipated because they will provide key insights into
how such processes, that dictate the evolution of the galaxy
population, evolve over time.


\section*{Acknowledgements}
We are grateful to the anonymous referee for numerous insightful
comments that improved the clarity of the original manuscript. The
work presented in this paper would not have been possible without
the Garching SDSS catalog. SK is grateful to Christy Tremonti for
many useful discussions regarding the extraction of spectroscopic
parameters used in this study. Sven De Rijcke, James Binney, Lisa
Young, Phillip Podsiadlowski and Sukyoung Yi are thanked for
constructive comments.

SK acknowledges a Leverhulme Early-Career Fellowship, a BIPAC
Fellowship and a Research Fellowship from Worcester College,
Oxford. Part of the research presented in this study used the
undergraduate computing facilities in the Department of Physics at
Oxford.

Funding for the SDSS and SDSS-II has been provided by the Alfred
P. Sloan Foundation, the Participating Institutions, the National
Science Foundation, the U.S. Department of Energy, the National
Aeronautics and Space Administration, the Japanese Monbukagakusho,
the Max Planck Society, and the Higher Education Funding Council
for England. The SDSS Web Site is http://www.sdss.org/.

The SDSS is managed by the Astrophysical Research Consortium for
the Participating Institutions. The Participating Institutions are
the American Museum of Natural History, Astrophysical Institute
Potsdam, University of Basel, University of Cambridge, Case
Western Reserve University, University of Chicago, Drexel
University, Fermilab, the Institute for Advanced Study, the Japan
Participation Group, Johns Hopkins University, the Joint Institute
for Nuclear Astrophysics, the Kavli Institute for Particle
Astrophysics and Cosmology, the Korean Scientist Group, the
Chinese Academy of Sciences (LAMOST), Los Alamos National
Laboratory, the Max-Planck-Institute for Astronomy (MPIA), the
Max-Planck-Institute for Astrophysics (MPA), New Mexico State
University, Ohio State University, University of Pittsburgh,
University of Portsmouth, Princeton University, the United States
Naval Observatory, and the University of Washington.




\nocite{Kaviraj2006a}
\nocite{Kaviraj2006b}
\nocite{Kauffmann1}
\nocite{Kauffmann2}
\nocite{Brinchmann2004}
\nocite{Martin2005}
\nocite{Blake2004}
\nocite{Bernardi2003}
\nocite{Yi2005}
\nocite{Goto2003}


\bibliographystyle{mn2e}
\bibliography{references}


\end{document}